\documentclass[aps]{revtex4}
\usepackage{graphicx}% Include figure files
\def\be {\begin{equation}}
\def\ee {\end{equation}}
\def\ba {\begin{eqnarray}}
\def\ea {\end{eqnarray}}
\def\nn {\nonumber}
\def\a  {\alpha}
\def\b  {\beta}
\def\c  {\gamma}

\def\d  {\delta}

\def\e  {\epsilon}

\def\k  {\kappa}

\def\L  {\Lambda}
\def\m  {\mu}

\def\o  {\omega}
\def\O  {\Omega}
\def\p  {\pi}
\def\P  {\Pi}

\def\la {\label}
\def\le {\left}
\def\ri {\right}
\def\pa {\partial}
\def\f {\frac}

\def\no {\noindent}
\def\bi {\begin{itemize}}
\def\ei {\end{itemize}}

\def\vs {\vspace}

\def\ul {\underline}

\def\laq{\hbox{~}\raise 0.4ex\hbox{$<$}\kern -0.8em\lower 0.62ex\hbox{$\sim$}\hbox{~}}
\def\gaq{\hbox{~}\raise 0.4ex\hbox{$>$}\kern -0.7em\lower 0.62ex\hbox{$\sim$}\hbox{~}}
\begin{document}

\title{Spectrum of rotating black holes and its implications for Hawking radiation}

\author{Saurya Das}
\affiliation{Dept. of Physics, University of Lethbridge,\\
4401 University Drive, Lethbridge, Alberta T1K 3M4, CANADA}
\email{saurya.das@uleth.ca}

\author{Himan Mukhopadhyay and P. Ramadevi}
\affiliation{Dept. of Physics, Indian Institute of Technology, Bombay\\
Mumbai - 400 076, INDIA}
\email{himan@phys.iitb.ac.in,himan@iucaa.ernet.in,ramadevi@phy.iitb.ac.in}

\begin{abstract}
The reduced phase space formalism for quantising black holes has
recently been extended to 
find the area and angular momentum spectra of four dimensional
Kerr black holes. We extend this further
to rotating black holes in all spacetime dimensions and 
show that although as in four dimensions the
spectrum is discrete, it is not equispaced in general. 
As a result, Hawking radiation spectra from these black holes
are continuous, as opposed to the discrete spectrum predicted
for four dimensional black holes. 
\end{abstract}

\maketitle

\section{Introduction}

Several different approaches have shown by now that the quantum mechanical 
spectra of black hole observables (such as area, charge and angular momentum)
are discrete \cite{others,bdk,qnm}. Among them, the reduced
phase space quantisation technique of Barvinsky et al \cite{bdk}
confirms an earlier conjecture by Bekenstein and collaborators that the 
area spectrum of quantum black holes is indeed discrete as well as
{\it equispaced} \cite{bm}. The last property makes a rather 
interesting prediction that Hawking radiation from black holes will have a 
noticeably discrete spectrum, which can have interesting experimental consequences
\cite{bcr}. 
The results were extended to the case of charged black holes, 
which yielded a separate discrete spectrum for the electric charge.
The area spectrum was slightly more complicated in this case, 
although the uniform spacing property remained intact. 
These spectra were re-derived from algebraic approaches in \cite{bek2,dry}.

Recently the above formalism was extended by Gour and Medved to 
rotating (uncharged) black holes in four spacetime dimensions \cite{gm}
(and to rotating charged black holes in \cite{gm2}).  
They showed that the corresponding area spectrum is also discrete and
equispaced: $A = 8\p \le(n + m +1/2 \ri)$ where $n$ and $m$ are 
non-negative integers. While $n$ signifies the departure of the 
black hole from extremality, $m$ measures the classical angular momentum
of the black hole. In this paper, we extend their formalism to
include $3$-dimensional $BTZ$ black holes as well as $5$ and higher
dimensional Myers-Perry type rotating black holes with multiple angular momenta 
parameters. We show that while the area spectrum is 
discrete in each case, in general it is not equispaced. In particular, 
the $3$ and $5$ dimensional black holes have a non-uniform area spectrum. 
We also show that this makes Hawking radiation from these black holes 
continuous, as opposed to the 
discrete radiation spectrum characteristic of black holes of uniformly 
spaced area. 

This paper is organised as follows. In the next section, we review
some thermodynamic properties of the $BTZ$ black hole and derive the
quantum mechanical spectra of its area and angular momentum. In
section (\ref{5sec}), we consider the $5$-dimensional rotating black hole
with $2$ angular momentum parameters and derive the corresponding
spectra. Then in section (\ref{6sec}), we derive the area spectrum for 
six and higher dimensional black holes, whose area spectum is shown to be quite 
distinct from black holes in lower dimensions.  
In section (\ref{hrsec}), we study the implications of the area spectrum
that we found for Hawking radiation, and show that in most cases the 
radiation spectrum is a continuum, {\it unlike} that for four dimensional
black holes. 
We conclude with a summary, some open questions as well as some 
observations regarding adiabaticc invariants in quantum gravity 
in section (\ref{conclsec}).

\section{BTZ Black Holes}

First, let us consider a BTZ black hole, which is a solution of 
three dimensional Einstein equations with a negative cosmological 
constant $\L \equiv -1/\ell^2$. 
Its horizon radii, area, entropy, Hawking temperature, angular velocity and 
the first law of black hole mechanics that it satisfies are
given by \cite{btz} 
\ba
r_\pm &=& 
\ell \le[ 
\f{M}{2} 
\le( 1 \pm \sqrt{1 - \le( \f{J}{M\ell}  \ri)^2} \ri) \ri]^{1/2} \la{btzr+} \\
A &=& 2\p r_+ \la{btzarea} \\
%M &=& \f{A^2}{\ell\sqrt{2(A\p)^2 - (J\p^2)^2 }} \\
S_{BH} &=& \f{A}{4G_3} = 4\p r_+ \\
T_H &=&  \f{\k}{2\p} = \f{r_+^2 - r_-^2}{2\p \ell^2 r_+} \\
%\k &=& \f{16\p M \le[ A^2 - (J\p)^2 \ri]}{A\le[2A^2 - (J\p)^2 \ri]} \\
\O &=& \f{J}{2 r_+^2} \la{btzomega} \\
%\O&=& \f{MJ\p^2}{2A^2 - (J\p)^2} \\
dM &=& T_H~dS_{BH} + \O~dJ~, \la{1st} 
\ea
where $\k$ is the surface gravity at the horizon and 
following \cite{btz} we have assumed that the
three dimensional Newton's constant $G_3=1/8$. 
In the extremal limit $r_+ = r_-$ one has:
\be
M \ell = J~,~\le( r_+ \ri)_{extr} = \ell \sqrt{\f{M}{2}} 
= \sqrt{\f{J}{2\ell}}~~,~
A_{extr}  
= \p \sqrt{\f{2 J}{\ell}}~~,~~
S_{extr} 
= 2 \p \sqrt{\f{2 J}{\ell}}~,
\la{btzextr}
\ee
where we have assumed $J\geq 0$ without loss of generality. 

To derive quantum mechanical spectra, we start with the 
complete set of commuting observables $({M}, {J})$ 
and their conjugates $(\P_M,\P_J)$. 
Since $\P_M$ has the interpretation of difference between 
Schwarzschild times across a spacelike slice extending from the left to the
right wedge of a Kruskal diagram, following \cite{bdk,gm}, we impose the
periodicity condition:
\be
\P_M \sim \P_M + \f{2\p}{\k}~, 
\la{period1}
\ee
to incorporate the thermodynamic nature of the system under consideration. 
Next, we make the following transformation to a new set of phase space
variables $(X,P_X)$ which automatically incorporate the above periodicity:
\ba
X &=& \sqrt{\f{B(M,J)}{\p}}~\cos\le(\P_M \k\ri) \la{x1} \\
P_X &=& \sqrt{\f{B(M,J)}{\p}}~\sin \le(\P_M \k\ri)~,  \la{x2}
\ea 
where $B(M,J)$ is an hitherto unknown function. Computation of Poisson bracket yields: 
\be
\{X,P_X\} = \f{\k}{2\p} \f{\pa B}{\pa M} 
\ee
{}From (\ref{1st}) it follows that if the function $B(M,J)$ is chosen such
that:
\be
B(M,J) = S(M,J) + F(J) 
= 2A(M,J) + F(J) ~,
\ee
where $F(J)$ is arbitrary, then the above Poisson bracket becomes unity
and the transformation $(M,\P_M) \rightarrow (X,\P_X)$ is indeed 
canonical. With this choice, squaring and adding (\ref{x1}) and (\ref{x2}), we get:
\be
X^2 + P_X^2  = \f{2A(M,J) + F(J) }{\p}~.
\ee
Noting that for a given $J$, since the horizon area $A$ cannot be less than 
$A_{extr}$ given in (\ref{btzextr}), it follows that $\forall~J \geq 0$,
the $(X,P_X)$ phase space has a `hole' centred 
at the origin and with radius $\mbox{Min}\le(2A(M,J)+F(J)\ri) 
= 2A_{extr}(J) + F(J) = 2 \p \sqrt{2J/2\ell} + F(J) $ removed from it. 
Although this gives rise to quantisation ambiguities, these can  
be eliminated by the following unique choice of $F$:
\be
F(J) = 
- 2 \p \sqrt{\f{2 J}{\ell}} 
~,
\ee
such that: 
\be
B(M,J) = 2A(M,J) - 2 \p \sqrt{\f{2 J}{\ell}} 
\ee
and
\be
X^2 + P_X^2  = \f{2A(M,J)}{\p} - 2  \sqrt{\f{2 J}{\ell}}~.
\la{sho1}
\ee
Quantisation is now straightforward with the replacements:
$X \rightarrow {\hat X}$ and $P_X \rightarrow {\hat P_X} = - i\pa/\pa X$.
The LHS of (\ref{sho1}) is recognisable as the Hamiltonian of a 
harmonic oscillator of mass$=1/2$ and angular frequency $=2$. 
This results in the following spectra for entropy and horizon area:
\be
%S - 2 \p \sqrt{\f{2 J}{\ell}}
%&=&  2\p \le(n + \f{1}{2} \ri)~~,~n=0,1,2,\dots  \\
A -   \p \sqrt{\f{2J}{\ell}} =   \p \le(n + \f{1}{2} \ri)~.  
\ee
Next, the quantisation of angular momentum follows from the usual 
identification in $3$-dimensions: ${\hat J} = -i \pa/\pa\phi$, 
resulting in eigenfunctions: $\psi_m = \exp(i m\phi)$ with eigenvalues
 $m=0,1,2,\dots$. Thus the final spectrum of area is:
\be
A = \p\le( n + \sqrt{\f{2m}{\ell}} + \f{1}{2} \ri)~,~~n,m=0,1,2,\dots
\la{btzspec1}
\ee
where it is understood that the area above (and in subsequent sections) 
is measured in Planck units. 
Note that it is discrete, but {\it not} equispaced. 
It will be shown in section (\ref{hrsec}) that this will have important 
consequences for Hawking radiation. 
Also note that $m=0$ (no rotation) reduces the above spectrum to the 
equispaced spectrum conjectured in \cite{bm}. However, unlike the
non-rotating case, 
or those that result from the recent identification of black 
hole quasi-normal modes with Hawking radiation frequencies (see e.g. 
\cite{qnm1,qnm}), Eq.(\ref{btzspec1}) does not have the 
multiplicative factor of $\ln(2)$ or $\ln(3)$ in front. Moreover, 
our `ground-state area' is non-vanishing, suggesting that the 
the end stage of Hawking radiation is a Planck sized remnant.  
The above two features are generic for black holes in 
arbitrary spacetime dimensions, as will be seen in subsequent sections.

\section{Five-dimensional rotating black holes}
\la{5sec}

We now extend our results to five-dimensional rotating black holes. 
As suggested from the approaches in \cite{bdk,gm} and in the previous section,
the first step towards quantizing black hole horizon area 
is to find a set of mutually commutating operators and their 
conjugates. A naive extension of the analysis in the preceeding 
section fails for $d \geq 4$, 
since the different components of angular momenta
fail to commute. However, 
as shown in \cite{gm} this problem can be readily
cured by replacing the usual angular momentum 
components by the Euler components of 
angular momentum, which mutually commute \cite{LL}. 
Here we extend that analysis 
to $d=5$ black holes which has {\it two} rotation parameters \cite{mp}. 
The key observation that makes this possible is that the 
relevant rotation group $SO(4)$ is isomorphic to $SU(2) \times SU(2)$. 
Defining the rotation and Lorentz generators in $5$-spacetime dimensions: 
\begin{equation}
\hat{L}_i = \frac{1}{2} \epsilon_{ijk} \hat{M}_{jk}, \qquad \hat{K}_i = \hat{M}_{i4} \qquad i,j,k = 1, 2, 3
\end{equation}
where,
\begin{equation}
\hat{M}_{\mu \nu} = 
- i(x_\mu \partial_\nu -  x_\nu \partial_\mu) \qquad \mu, \nu = 1, 2, 3, 4 ~,
\end{equation}
their commutation relations follow:
\begin{equation}
[\hat{L}_i , \hat{L}_j ] = i \epsilon_{ijk} \hat{L}_k , \qquad 
[\hat{L}_i , \hat{K}_j ] = i \epsilon_{ijk} \hat{K}_k , \qquad 
[\hat{K}_i , \hat{K}_j ] = i \epsilon_{ijk} \hat{L}_k \; .
\end{equation}
Next, defining two new angular momentum operators $\hat{J}_1$ and $\hat{J}_2$ as
\begin{eqnarray}
\hat{J}_{1i} = \frac{1}{2} (\hat{L}_i + \hat{K}_i) \; , \\
\hat{J}_{2i} = \frac{1}{2} (\hat{L}_i - \hat{K}_i ) \; ,
\end{eqnarray}
it can be shown that
\begin{equation}
[\hat{J}_{1i} , \hat{J}_{1j}] = i \epsilon_{ijk} \hat{J}_{1k}, \qquad 
[\hat{J}_{2i} , \hat{J}_{2j}] = i \epsilon_{ijk} \hat{J}_{2k}, \qquad
[\hat{J}_{1i} , \hat{J}_{2j}] = 0 \; .
\end{equation}
giving two copies of the $SU(2)$ algebra. 
Accordingly, defining two copies of the analogs of 
Euler components of the angular momenta,
$J_{\a i}, J_{\b i}, J_{\gamma i}, (i=1,2)$, as follows:
\begin{eqnarray}
J_{11} & = & -\cos \alpha_1 \cot \beta_1 J_{\alpha_1} - \sin \alpha_1 J_{\beta_1} + \frac{\cos \alpha_1}{\sin \beta_1} J_{\gamma_1} \; ,\la{j11} \\
J_{12} & = & - \sin \alpha_1 \cot \beta_1 J_{\alpha_1} + \cos \alpha_1 J_{\beta_1} + \frac{\sin \alpha_1}{\sin \beta_1} J_{\gamma_1} \; ,\\
J_{13} & = & J_{\alpha_1}  \; ,\\
J_{21} & = & -\cos \alpha_2 \cot \beta_2 J_{\alpha_2} - \sin \alpha_2 J_{\beta_2} + \frac{\cos \alpha_2}{\sin \beta_2} J_{\gamma_2} \; ,\\
J_{22} & = & - \sin \alpha_2 \cot \beta_2 J_{\alpha_2} + \cos \alpha_2 J_{\beta_2} + \frac{\sin \alpha_2}{\sin \beta_2} J_{\gamma_2} \; , \\
J_{23} & = & J_{\alpha_2} \; , \la{j23}
\end{eqnarray}
where the six coordinates 
$\a_i,\b_i,\c_i,~i=1,2$ are $5$-dimensional analogues of Euler angles in four dimensions. 
Finally, we define the two `classical angular momenta':
\begin{eqnarray}
J_{1Cl}^2 = J_{11}^2 + J_{12}^2 + J_{13}^2 = \frac{1}
{\sin^2 \beta_1}
[J_{\alpha_1}^2 + J_{\gamma_1}^2 - 2 \cos \beta_1 J_{\alpha_1} J_{\gamma_1} ] + J_{\beta_1}^2 \la{eu5.1} \\
J_{2Cl}^2 = J_{21}^2 + J_{22}^2 + J_{23}^2 = \frac{1}
{\sin^2 \beta_2}
[J_{\alpha_2}^2 + J_{\gamma_2}^2 - 2 \cos \beta_2 J_{\alpha_2} J_{\gamma_2} ] + J_{\beta_2}^2  . \la{eu5.2}
\end{eqnarray}
%So, $J_{1Cl}$ and $J_{2Cl}$ are the two Casimir operators. 
%
Now the above two angular momenta are related to the ones that 
enter the five dimensional black hole metric in the following way \cite{hhtr}:
\ba
J_\phi &=& \f12 \le( J_{1Cl} + J_{2Cl} \ri) \la{jphi} \\
J_\psi &=& \f12 \le( J_{1Cl} - J_{2Cl} \ri)  \la{jpsi} 
\ea
which in turn defines the parameters $a$ and $b$ as:
\ba
J_\phi &=& \f23 M a \\
J_\psi &=& \f23 M b ~,
\ea
implying:
\ba
J_{1Cl} &=& \f23 M \le( a + b \ri) \\
J_{2Cl} &=& \f23 M \le( a - b \ri)~. 
\ea
The two horizons are roots of the equation
%
%\endnote{Note that $r$ here is not the usual radial coordinate. In 4+1 dimensions
%it is defined by $\sum_{i=1}^2 \frac{x_i^2 + y_i^2}{r^2 + a_i^2} = 1$.}:
%
\begin{equation}
(r^2 + a^2)(r^2 + b^2) - \mu r^2 = 0 \; .
\end{equation}
The resulting thermodynamic quantities are (with $M \equiv \frac{3 \pi}{8} \mu$):
\ba
2 r_\pm^2 &=& \m - a^2 - b^2 \pm \sqrt{(\m-a^2-b^2)^2-4a^2b^2} \la{hor5.1}\\
A &=& 
\f{2\p^2 \m}{\k}\le[  1 
- \f{a^2}{r_+^2 + a^2}  - \f{b^2}{r_+^2 + b^2}   
\ri] 
\la{a5.2} \\
%A &=& 2\p^2 r_+^3 \\
%
S_{BH} &=& \f{A}{4G_5} \\
T_H &=& \f{\k}{2\p} = \f{2r_+^2 +  a^2 + b^2  - \m}{2\p \m r_+} \la{arb1}\\
\O_\phi &=& \f{a}{r_+^2 + a^2} ~ \la{5domegas1} \\
\O_\psi &=& \f{a}{r_+^2 + b^2} ~ \la{5domegas2} \\
dM &=& T_H~dS_{BH} + \O_\phi~dJ_{\phi} + \O_\psi~dJ_{\psi}~, \\
\ea
{}From (\ref{hor5.1}) the extremality bound follows: 
\begin{eqnarray}
\nonumber
\mu & \geq & a^2 + b^2 + 2 | a b | \qquad \hbox{or equivalently},\\
M^3 & \geq & \frac{27 \pi}{32} (J_{\phi}^2 + J_{\psi}^2 + 2 | J_{\phi} J_{\psi} |)  \; .
\end{eqnarray}
%
%which when saturated gives: 
%
%\begin{equation}
%r_{ext}^2 = | a b | = \frac{9}{4 M^2} |J_{\phi} J_{\psi} | \; .
%\la{rext5}
%\end{equation}
%%%
{}From (\ref{hor5.1}), (\ref{a5.2}) and (\ref{arb1}) it follows that:
\ba 
A &=& 
%{(r_+^2 + a^2)(r_+^2 + b^2)}
\f{2\p^2 \m^2 r_+}{2r_+^2 + a^2 + b^2 - \m}
~\f{r_+^4 - (ab)^2}{(r_+^2 + a^2)(r_+^2 + b^2)}~,
%\f{\sqrt{\le( \m - a^2 - b^2\ri)^2 - (2ab)^2}}{2\p \m r_+}~,  
%\f{\p^2 \m^2 r_+^2 \le( r_+^2 + ab \ri) }{(r_+^2 + a^2)(r_+^2 + b^2)} 
\la{arb2} \\
T_H &=& 
%\f{2\p^2 \m^2 r_+}{2r_+^2 + a^2 + b^2 - \m}
%{(r_+^2 + a^2)(r_+^2 + b^2)}
%\f{r_+^4 - (ab)^2}{(r_+^2 + a^2)(r_+^2 + b^2)}~,
\f{\sqrt{\le( \m - a^2 - b^2\ri)^2 - (2ab)^2}}{2\p \m r_+}~,  
\ea
which along with Eq.(\ref{hor5.1}) 
implies the following conditions in the extremal limit 
($T_H=0$):
\ba
\mu &=& (a+b)^2 \\
r_{ext}^2 &=& | a b | = \frac{9}{4 M^2} |J_{\phi} J_{\psi} | \; .
\ea
Thus, from (\ref{arb2}) it follows that 
that the horizon area in the extremal limit is:
\begin{equation}
A_{ext} = 8 \pi \sqrt{J_{\phi}J_{\psi}} 
= 4\p \sqrt{|J_{1Cl}^2 - J_{2Cl}^2 | }  = A_{ext}\le( J_{1Cl}, J_{2Cl} \ri) \; .\la{aext}
\end{equation}
Now, it can be seen from (\ref{eu5.1})and (\ref{eu5.2}) that $J_{iCl}^2$ commutes
with $J_{\a i}$ and $J_{\gamma i}$, for $i=1,2$. Thus, one can choose as 
canonical variables the following set: 
$\le(M, J_{1Cl}, J_{\alpha_1}, J_{\gamma_1},
J_{2Cl}, J_{\alpha_2},J_{\gamma_2}\ri)$ with the corresponding 
conjugates $\le(\Pi_M, \Pi_{J_1}, \Pi_{\alpha_1}, \Pi_{\gamma_1},
\Pi_{J_2},\Pi_{\alpha_2},\Pi_{\gamma_2}\ri)$ respectively.
As before, we impose the periodicity (\ref{period1}), which is now
incorporated by the following transformation:
\begin{eqnarray}
X =\sqrt{ \frac{B(M, J_{1Cl} , J_{\alpha_1} , 
J_{\gamma_1}, J_{2Cl}, J_{\alpha_2}, 
J_{\gamma_2})}{\pi}} \cos(\Pi_M \kappa) \; ,\la{ct5.1} \\
P_X = \sqrt{ \frac{B(M, J_{1Cl} , J_{\alpha_1} , 
J_{\gamma_1}, J_{2Cl}, J_{\alpha_2}, 
J_{\gamma_2})}{\pi}} \sin(\Pi_M \kappa) \; .\la{ct5.2}
\end{eqnarray}
Computation of Poisson brackets yields once again: 
\be
\{X,P_X\} = \f{\k}{2\p} \f{\pa B}{\pa M}~,
\ee
which now implies:   
%
%
%Demanding that it is a canonical transformation, from the fact that $[X, P_X]_{PB} = 1$, it turns out,
%as in the previous case that $\frac{\partial A}{\partial M} = 4 \frac{\partial B}{\partial M}$. So,
%now, the new set of canonical variables consists of $X, J_{1Cl} , J_{\alpha_1} , J_{\gamma_1}, J_{2Cl}, J_{\alpha_2}$ and
%$ J_{\gamma_2}$ and the corresponding canonical conjugates are $P_X, P_{1Cl}, P_{\alpha_1}, P_{\gamma_1},
%P_{2Cl}, P_{\alpha_2}$ and $P_{\gamma_2}$.
%
%The above relationship implies
%\begin{equation}
%B(M , J_{1Cl} , J_{2Cl}, J_{\alpha_1} , J_{\gamma_1}, J_{\alpha_2}, J_{\gamma_2}) = \frac{1}
%{4} A(M , J_{1Cl}, J_{2Cl} ) + F (J_{1Cl} J_{2Cl}, J_{\alpha_1} , J_{\gamma_1}, J_{\alpha_2}, J_{\gamma_2} ) \; .
%\end{equation} 
%Choosing $F = A_{ext}$
\begin{equation}
B = \frac{1}{4}[ A(M , J_{1Cl}, J_{2Cl} ) - A_{ext}(J_{1Cl},J_{2Cl})] \; ,
\end{equation}
where we have set $G_5=1$. 
Squaring and adding (\ref{ct5.1}) and (\ref{ct5.2}): 
\begin{equation}
X^2 + P_X^2 = \frac{1}{4 \pi}[ A(M , J_{1Cl}, J_{2Cl} ) - A_{ext}(J_{1Cl},J_{2Cl})] 
\geq 0 \; .
\la{sho51}
\end{equation}
%%which, in operator language, can be expressed as follows 
%\begin{equation}
%\frac{1}{2 \pi} \hat{B} = \frac{1}{8 \pi}[ A(M , J_{1Cl}, J_{2Cl} ) - A_{ext}] = \frac{\hat{X}^2}{2}
%+ \frac{\hat{P_X}^2}{2} \; .
%\end{equation}
%Therefore, the eigenvalues of the operator $\hat{B}$ are the same as in the case of a 3+1 
%dimensional black hole, namely,
%\begin{equation}
%B_n = 2 \pi \hbar \left( n + \frac{1}{2} \right) \; .
%\end{equation}
%So, the difference of area spectrum of a 4+1 dimensional black hole from that of a 3+1 
%dimensional one arises solely due to the different quantization of the extremal area, which s 
%to be found out next. 
%
%
Quantising as before, we now get:
\be
A - A_{ext} = 8\p \le( n + \f{1}{2} \ri)~,~n=0,1,2,\cdots
\ee
It may be noted that the above equation has the same form as the corresponding 
equation in the $d=4$ case (Eq.(43) of \cite{gm}). The distinction between
that case and the current one lies in the difference in the 
functional form of extremal entropy and the resulting difference that arises
on quantisation. Now, since (\ref{j11})-(\ref{j23}) remains valid when 
the classical quantities are replaced by their corresponding operator
counterparts \cite{LL,gm}, together with the replacement:
$ \hat J_{ai} \rightarrow -i\f{\pa}{\pa a_i}~,a=\a,\b,\gamma$, one obtains
the following relations for $i=1,2$:
%
%
%The extremal area depends on $J_{1Cl}$ and $J_{2Cl}$. 
%The eigenvalues of the corresponding 
%operators can be found from the following considerations.
%
%Clearly,
\begin{eqnarray}
\hat{J}_{iCl}^2 &=& \frac{1}{\sin^2 \beta_i}
[\hat{J}_{\alpha_i}^2 + \hat{J}_{\gamma_i}^2 - 2 \cos \beta_i 
{\hat J}_{\alpha_i} {\hat J}_{\gamma_i} ] + {\hat J }_{\beta_i}^2 \; , \la{ang4} \\
\hat{J}_i^2 &=& \hat{J}_{i1}^2 + \hat{J}_{i2}^2 + \hat{J}_{i3}^2 \; .\\
\mbox{and}~~\hat{J}_{iCl}^2 - \hat{J}_i^2 &=& 
\cot \beta_i \frac{\partial}{\partial \beta_i}~. 
\la{j4}
\ea
Since all the operators in the set
$\{\hat{J}_1, \hat{J}_2, \hat{J}_{1Cl}$ and $\hat{J}_{2Cl}\}$ commute
with $\hat{J}_{\alpha_i}$ and $\hat{J}_{\gamma_i}~(i=1,2)$, both 
sets of eigenvectors 
$\{|j_1, j_2, J_{\alpha_1}, J_{\alpha_2}, J_{\gamma_1}, J_{\gamma_2} \rangle\}$ 
and 
$\{|J_{1Cl}, J_{2Cl}, J_{\alpha_1}, J_{\alpha_2}, J_{\gamma_1}, J_{\gamma_2} 
\rangle\}$ are complete and an element of one can be expressed as a 
superposition of the elements of the other: 
%
%
%there are four sets of eigenbasis, among which one is $|j_1, j_2, J_{\alpha_1},
%J_{\alpha_2}, J_{\gamma_1}, J_{\gamma_2} \rangle$ and another is 
%$|J_{1Cl}, J_{2Cl}, J_{\alpha_1}, J_{\alpha_2}, J_{\gamma_1}, J_{\gamma_2} 
%\rangle$. 
%So,
\begin{equation}
|J_{1Cl}, J_{2Cl}, J_{\alpha_1}, J_{\alpha_2}, J_{\gamma_1}, J_{\gamma_2} \rangle
 = \sum_{j_1, j_2} 
C_{j_1, j_2, J_{1Cl}, J_{2Cl}} \le( j_1, j_2, J_{1Cl}, J_{2Cl} \ri)  
|j_1, j_2, J_{\alpha_1},
J_{\alpha_2}, J_{\gamma_1}, J_{\gamma_2} \rangle \; ,
\la{super1}
\end{equation}
%where, $C_{j_1, j_2, J_{1Cl}, J_{2Cl}}$ is a complex number depending upon
%only $j_1, j_2, J_{1Cl}$ and $J_{2Cl}$.
%
To compute the eigenvalues of $J_{iCl}$, consider the 
eigenfunction with zero eigenvalues for $\a_i$ and $\gamma_i$:
$\Psi_{J_{1Cl},0,0,0,0,0} = \langle \alpha_1, \beta_1,
\gamma_1, \alpha_2, \beta_2, \gamma_2 | J_{1Cl}, 0, 0, 0, 0, 0 \rangle$.
Then, from (\ref{ang4}):
\begin{equation}
\hat{J}_{iCl}^2 \Psi_{J_{iCl},0,0,0,0,0} = \hat{J}_{\beta_i}^2 \Psi_{J_{iCl},0,0,0,0,0} = - 
\frac{\partial^2}{\partial \beta_1^2} 
\Psi_{J_{iCl},0,0,0,0,0} \equiv  m_{i}^2~\Psi_{J_{iCl},0,0,0,0,0} \; .
\end{equation}
With the usual identification for Euler angles, 
$\beta_i + \pi = \pi - \beta_i$, we get:
\begin{eqnarray}
\nonumber
\Psi_{J_{iCl},0,0,0,0,0} & \sim & \cos \le(m_{i} \beta_i \ri)~,~~  
m_{i} = 0,1,2.\cdots \la{5dquant1} \\
& \sim & \sin \le(m_{i} \beta_i \ri) ~,~~ m_{i} = 1/2, 3/2, 5/2, \cdots \\
\mbox{with}~~ 
J_{iCl} &=& m_i ~.
\end{eqnarray}
The additional fact that $\Psi_{j_i,0,0,0,0}$ is a symmetric function of $\b$ 
together with equation (\ref{super1}) rules out the half-odd integral 
quantisation of $J_{iCl}$. Thus we are left with Eq(\ref{5dquant1}) 
as the correct quantisation condition. This implies, from (\ref{jphi}) 
and (\ref{jpsi}) that:
\ba
J_\phi &=& \f12 \le( m_1 + m_2 \ri) \\
J_\psi &=& \f12 \le( m_1 - m_2 \ri)~,~ 
m_{i} = 0,1,2.\cdots 
\ea
Using (\ref{aext}), the final result of the area spectrum is 
\begin{equation}
A =  8 \pi \left( n + \f12 \sqrt{|m_1^2 - m_2^2|} + \frac{1}{2} \right)~,~~
n, m_{1}, m_{2} = 0,1,2,\cdots
\la{5dspec2}
\end{equation}
Thus we see that the five-dimensional Kerr black hole area spectrum
is also quantized but not equally spaced.
%
%Note that as for the $d=3,4$, 
%the area spectrum is quantized; however as in $d=3$, it is 
%not equally spaced in general. 
Equispaced spectrum can arise as a special case however, either when 
$m_1=0$ (or $m_2=0$) or when $m_1=m_2$. 
The latter case implies $J_\psi=0$, corresponding to a single parameter
five dimensional black hole. 
We will examine in section (\ref{hrsec}) the effects of 
quantized area spectrum on Hawking radiation. 
%We will examine these cases separately in 
%section (\ref{hrsec}) to see its effects on Hawking radiation. 

\section{Six and higher dimensional rotating black holes} 
\la{6sec}

The situation simplifies drastically for six and higher dimensions because of 
the absence of an extremality bound. The number of angular momentum parameters
$a_i$'s for the $d$-dimensional Kerr black hole will be 
$[(d-1)/2]$(square bracket denotes integer part).
For simplicity, we consider 
the metric component of a  single angular momentum 
parameter Kerr black hole- that is,  $a_1\equiv a \equiv 2J/M$
where $M= \mu (d-2)(2\pi)^{(d-1)/2}/  
\{\Gamma[(d-1)/2] 16 \pi G\}$ and 
$a_i=0$ for $2 \leq i  \leq [(d-1)/2]$ \cite{mp}: 
%the absence of an extremality bound. In other words, the black hole angular
%momentum is in no way contrained by its mass and in fact can be arbitrarily large
%\cite{mp}. This can be most easily seen for rotating black hole with one 
%angular momentum parameter $a$, for which:
%
\be
g^{rr} = \f{r^{d-5} \le( r^2+a^2 \ri)-\m}{r^{d-5} \le(r^2 + a^2 \cos^2 \theta \ri)}~.
\ee
It follows that for $d\geq 6$, the horizon condition $g^{rr}=0$ is satisfied for some 
$0 \leq r \leq \infty$ for any given value of $a$ and $M$. 
In other words, the black hole angular
momentum is in no way contrained by its mass and in fact can be arbitrarily large.
Moreover, for a fixed $a$, for $\m \ll a$, we see that the horizon radius 
shrinks to zero. This can also be seen from Fig.1, where the outer 
horizon radius has been 
plotted against the black hole mass for $6 \leq d\leq 10$
Thus, there is no `minimum' horizon area for a fixed
$a$, and the function $F(J)$ can be chosen to vanish. Consequently, 
the equivalents of (\ref{ct5.1})-(\ref{ct5.2}) read as:
\begin{figure}\label{figure1}
\begin{center}
\includegraphics[scale=.33, angle=270]{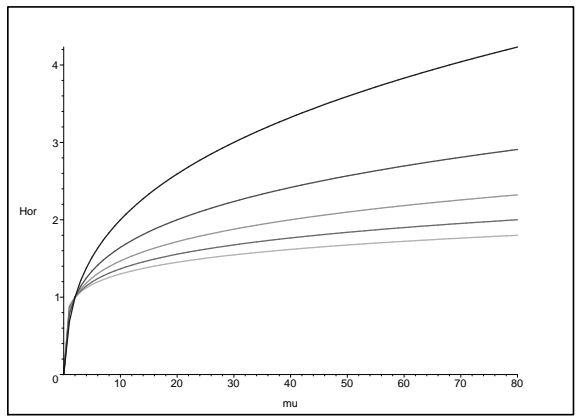}
\caption{
Plot of $r_+$ vs $\m$ for $d=6,7,\dots,10$ for $a=1$. The highest graph
corresponds to $d=6$ and the lowest to $d=10$.  }
\end{center}
\end{figure}
\begin{eqnarray}
X =\sqrt{ \frac{A(M,J)}{4 \pi}}~\cos(\Pi_M \kappa) \; ,\la{ct6.1} \\
P_X = \sqrt{ \frac{A(M,J)}{4 \pi}}~\sin(\Pi_M \kappa) \; .\la{ct6.2}
\end{eqnarray}
where
\ba
\k &=& \f{\pa \P/\pa r - 2\m r}{2\m r^2}|_{r_+}~,~~d=\mbox{odd} \\ 
   &=& \f{\pa \P/\pa r - \m}{2\m r}|_{r_+}~,~~~~~d=\mbox{even}  
\ea
and
$\P = \prod_{i=1}^{[\f{d-1}{2}]} \le( r^2 + a_i^2 \ri)$. 
%$[x]$ denoting the greatest integer contained in $x$.  
Once again (\ref{ct6.1})-(\ref{ct6.2}) incorporate the natural periodicity 
condition (\ref{period1}) and the transformations are canonical. 
Squaring and adding:
\be
A = 4\p \le( X^2 + P_X^2 \ri)~.
\la{sho61}
\ee
Quantisation yields:
\be
A_n = 8\p \le( n + \f{1}{2} \ri) ~,~~n=0,1,2,\cdots.
\la{area6}
\ee
Remarkably, angular momentum plays no role in area spectrum in $d \geq 6$, 
and the above spectrum is {\it identical} to the neutral black hole 
spectrum found in \cite{bdk}. The Hawking radiation spectrum is not the 
same however, as will be shown in the next section. It would be interesting to 
investigate other physical significances that this equispaced spectrum might
have.

\section{Hawking Radiation from Rotating Black Holes} 
\la{hrsec}

An important consequence of 
a strictly equispaced area spectrum is a markedly discrete
spectrum for Hawking radiation \cite{bm,bcr,bdk}. Consider for example the 
area spectrum of a $d=4$ Schwarzschild black hole.  
Using $A = 4\p r_+^2 = 16\p M^2$, the lowest Hawking radiation frequency $\o_0$
emitted due to a transition of 
this black hole from an excited state to its next lower state 
($\d n = -1,~ |\d A| = 8\p$ from (\ref{area6}) ) is given by:
\be
\o_0 = |\d M| = \f{|\d A|}{32\p M} = \f{1}{4M} = 2\p T_H 
\la{scfreq1}
\ee
and all higher frequencies will be multiples of the above. Comparing with 
Wien's law for frequency of maximum intensity of blackbody radiation:
\be
\o_{max} \approx T_H~, 
\ee
we see that the $\o_0 \approx \o_{max}$. Thus there will be only a few visible
spectrum lines in Hawking radiation {\it even for a macroscopic black hole}, 
leading to the aforementioned conclusion.   
The qualitative picture remains unchanged for Reissner-Nordstr\"om
black holes as well \cite{bdk}. 
Next we examine the impications for rotating black holes.

\subsection{BTZ black hole}

In this case, it follows from 
(\ref{btzspec1}) that for an emitted quantum of Hawking radiation:
\be
\d S_{BH} = 2\d A = 2\p \le[ \d n + \f{\d m}{\sqrt{2\ell m}} \ri]~.
\ee
Replacing $dM \rightarrow \o_0$ and plugging in the expression for 
$\O$ from (\ref{btzomega}) in the first law (\ref{1st}), we get:
\be
\o_0 = 2\p T_H~\d n + 
\le[ \f{2\p T_H}{\sqrt{2\ell m}} 
+ \f{m}{2\le(n + \sqrt{\f{2m}{\ell}} + 1/2  \ri)^2} \ri] \d m
\ee
%
%
%Using (\ref{btzr+})-(\ref{btzarea}) and the area spectrum (\ref{btzspec1}), 
%and repeating the preceeding steps, we find:
%%
%\be
%\o_0 = 2\p T_H \le[ |\d n| + \f{|\d m|}{\sqrt{2m}} \ri] ~.
%\ee 
%%
Next, consider an emission line for which $\d n = 0$ and $|\d m| = 1$, 
and the black hole to be `macroscopic', for which $n \geq m \gg 1$. Then:
\be
\o_0 = \f{2\p T_H}{\sqrt{2\ell m}} 
+ \f{m}{2\le(n + \sqrt{\f{2m}{\ell}} + 1/2  \ri)^2}  \rightarrow 0 
\ee
implying that the radiation spectrum is quasi-continuum, quite different 
from the $d=4$ case and exactly as predicted
by the original Hawking analysis. 
We will return to the case $\d n \neq 0$ for this and other black holes
in the concluding section.

\subsection{Four dimensional Kerr black hole} 

Here we use the spectrum found in \cite{gm}, namely:
\ba
A &=& 8\p \le( n + m + \f12\ri)~,~~n,m=0,1,2,\cdots~, \la{a10} \\ 
J_{Cl} &=& m \la{j10}
\ea
implying
\be
\d A = 8\p \le( \d n + \d m \ri)~,~~\d n,\d m=0,1,2,\cdots~. 
\ee
Once again, using (\ref{a10})-(\ref{j10}), 
the first law and $\O = a/(r_+^2 + a^2)$, we now get:
\be
\o_0 = 2\p T_H~\d n + \le[ 2\p T_H
+ \f{m}{\f{M}{2} \le( n + m + \f12 \ri) + \f{m^2}{M} }
\ri] \d m
\ee
In this case, for $|\d m| = 1$ (and for any finite $\d n$), we see that:
\be
\o_0 \approx 2\p T_H
\ee
and the Hawking radiation spectrum is once again discrete and 
%
%and the classical horizon area formula:
%%
%\be
%A = 4\p r_+^2 = 4\p \le( M + \sqrt{M^2 - a^2} \ri)^2~, 
%\ee
%%
%where $a=J/M$, to get:
%%
%\be
%\o_0 = 2\p T_H~\le( |\d n| + |\d m| \ri) + \f{a |\d m| }{r_+^2 + a^2} 
%%\approx 2\p T_H ~,
%\ee
%
identical to the Schwarzschild (and Reissner-Nordstr\"om) 
black hole spectrum (\ref{scfreq1}).

\subsection{Five dimensional rotating black hole}

From the five dimensional area spectrum (\ref{5dspec2}), 
the first law and the expression for the angular momenta 
(\ref{5domegas1}-\ref{5domegas2}), 
we get:
\be
\o_0 = 2\p T_H \le( |\d n| + 
\f{m_1 |\d m_1| - m_2 |\d m_2|}{2 \sqrt{m_1^2 - m_2^2}} 
%\f{m_1 |\d m_2| + m_2 |\d m_1|}{2\sqrt{m_1 m_2}} 
\ri)
+ \f{3}{8M} \le[
\f{(m_1+m_2) d(m_1 + m_2)}{r_+^2 + 9(m_1+m_2)^2/16M^2}
+ (m_2 \rightarrow - m_2) \ri]
%
%
%+ \sum_{i=1}^2 \f{3m_i/2M}{r_+^2 + 9m_i^2/4M^2}~\d m_i  ~.
\la{hr5dim1}
\ee
where we have assumed without loss of generality that $m_1 \geq m_2$. 
As for BTZ, consider an emission characterised by: $\d n = 0$ 
(signifying $A-A_{ext}$ remaining unchanged)  
and concentrate on the second term within the 
parenthesis, since the last term is in any case negligible for 
very large $M$. We see that for any given
$ \e \ll 1~,$
if 
\be
\d m_1 = \le( \f{m_2}{m_1} \ri) \le[ 1    
+ \f{2\e}{\d m _2} \sqrt{\le( \f{m_1}{m_2}\ri)^2 -1 } \ri] \d m_2~ \approx 
\le( \f{m_2}{m_1} \ri) \d m_2~,  
\la{5dhr1}
\ee
then 
\be
\o_0 = 2\p \e~T_H \rightarrow 0 
\ee
Notice however that both $|\d m_{1,2}| \geq 1$. This implies, 
along with (\ref{5dhr1}) and $|\d m_2| \leq m_2$, that $m_2 \leq m_1 \leq m_2^2$.  
That is, unless the two angular momentum parameters are hugely 
disproportionate, such that the second inequality in the above is
violated (or when $m_1=0$ or $m_2=0$), 
the radiation spectrum is a quasi-continuum, as was the case in 
three dimensions. 
Since these black holes are natural
candidates in certain brane world scenarios, it is possible that 
the above observations will have 
experimental signatures from the four dimensional (brane) point of
view. 
%The above conclusions must be accompanied by the following caveat: 
%as noted at the end of section (\ref{5sec}), 
%for the special cases $m_1=0,~m_2=0$ or $m_1=m_2$, the 
%area spectrum is equispaced and the Hawking spectrum is discrete. 
%This can also be seen from expression (\ref{hr5dim1}).

\subsection{Six and higher dimensional rotating black hole}

{}From (\ref{area6}) and the first law, we get:
\be
\o_0 = 2\p T_H~\d n + \f{a}{r_+^2 + a^2}~\d m ~.
\ee
Note that, as for $d=4$ Schwarzschild black holes, the first term 
implies a discrete spectrum. But unlike that case, here an
additional parameter $\d m$ ($\d m_i, i = 1,2, [d-1]/2$ in general) 
enters the problem. Recalling that there is no extremality bound for $d \geq 6$, 
if the black hole is such that $a \gg r_+$ and the
black hole to be macroscopic, then for $\d n = 0$, $|\d m|=1$, 
we have:
\be
\o_0 = \f{1}{a}~\rightarrow 0.
\ee
It is easy to see that similar conclusions will follow for 
$ a \ll r_+$ as well as $a \approx r_+$. 
Thus we see that the single angular momentum Kerr black holes
in $d \geq 6$ shows continuous Hawking spectrum in spite of
the area spectrum being equispaced. It remains to be checked whether
this property holds even if we include more angular momentum
parameters.

%Thus, 
%the Hawking spectrum is continuous in general, with a 
%word of caution: for certain special regions in the parameter space of these
%black holes, corresponding to one or more of the angular momentum parameters
%being equal, the spectrum may be discrete, as was seen for $d=5$. 

\section{Conclusions}
\la{conclsec}

In this paper, we have extended an earlier formalism by Barvinsky et al,
adapted to Kerr black holes by Gour and Medved to include rotating black holes
in all spacetime dimensions. This was done in three steps. 
First, the method was applied to $3$-dimensional $BTZ$ black holes with 
one angular momentum parameter, where the observables $A$ and $J$ were 
mutually commuting. Second, it was applied to $5$-dimensional
rotating black holes with two angular momentum parameters. Using the
fact that the corresponding rotation group $SO(4) \cong SU(2) \times SU(2)$, 
and using the technique of \cite{gm} in which the Cartesian components
of angular momentum are replaced by the mutually commuting Euler components,
we were able to quantise the the two angular momentum components and 
arrive at the spectrum of horizon area of these black holes. Finally six and
higher dimensional black holes were studied with one
angular momentum parameter, since this was sufficient to 
demonstrate its essential difference with lower dimensional black holes, 

The horizon area spectra for the above black holes
were found to be quite different from that in $d=4$.
In particular, we found that both for $d=3$ and $5$ the spectrum is
no longer equispaced 
(except for the special case when the two angular momentum 
parameters were equal in $d=5$). A direct consequence was that 
Hawking radiation spectrum from these black holes are practically continuous
(for large black holes), 
as opposed to the distinctly discrete spectrum predicted for 
$d=4$ black holes which can carry angular momentum or charge. 
The situation was even more interesting for $d=6$, in which case although 
the spectrum turned out to be equispaced, the Hawking radiation retained
its continuous nature. 
Although it may seem that there is an additional discrete spectrum for
all the cases considered when $\d n \approx 1$, this does not affect our 
conclusions in any way, since the superposition of a discrete and a continuous
spectrum is a continuum, the greybody factors being the same for each case. 
Such closeness of radiation spectra from higher 
dimensional black holes with the semi-classical Hawking analysis may lead one
to speculate 
that brane world black holes are more fundamental, since Hawking's result
is expected to be valid for macroscopic black holes. 
In fact one of the proposed brane world scenarios requires the 
spacetime dimension of our universe to be at least six \cite{add}, 
for which we have seen that 
interesting conclusions follow for rotating black holes. 
However, further investigations are required before arriving at 
a definitive conclusion. 
In any case, phenomenological 
implications of our higher dimensional results seem worth investigating. 

We conclude with an observation related to {\it adiabatic invariants}
in quantum gravity. Motivated by some thought experiments, it was 
conjectured by Bekenstein that black hole horizon area is an adiabatic 
invariant in quantum gravity, i.e. it remains unaffected by 
small changes in black hole parameters. In \cite{bdk} an explicit proof
of this conjecture was presented for spherically symmetric uncharged black holes,
while in the presence of a charge $Q$ it was shown that $A- A_{extr}(Q)$ was an adiabatic
invariant. We show here that a similar result follows. 
For periodic systems, it is well known that the quantity:
\be
{\cal I} \equiv \oint P_X dX 
\ee
is an adiabatic invariant, where $(X,\P_X)$
are usual phase-space coordinates \cite{mech}. It follws that for harmonic oscillator
like Hamiltonians as in (\ref{sho1}), (\ref{sho51}) and (\ref{sho61}), the corresponding 
adiabatic invariants are: 
\ba
{\cal I} &=& \f{A-A_{extr}(J)}{\p} ~~~~(BTZ)\nn \\ 
         &=& \f{A-A_{extr}(J_{1Cl},J_{2Cl})}{8\p} ~~~~(d=5)\nn \\ 
         &=& \f{A}{8\p} ~~~~( d \geq 6)~. \nn
\ea 
Note once again that as for charged black holes, upto $d=5$, the 
`excess-area' above extremality is an adiabatic invariant, while
for $d\geq 6$, the horizon area itself is the invariant, as 
conjectured by Bekenstein. It will be interesting to study the most
general case of charged and rotating black holes to see what the adiabatic
invariants are for such a system.

\vs{.4cm}
\no
\ul{{\bf Acknowledgements:}}

\vs{.2cm}
\no
We thank G. Kunstatter for useful comments. 
This work was supported in part by the
Natural Sciences and Engineering Research Council of Canada and funds of the
University of Lethbridge.

\end{document}